\documentclass[twocolumn]{aastex62}
\usepackage{float,graphicx,amsmath,multirow,mathtools}
\usepackage{color}
\usepackage[version=3]{mhchem}
\newcommand{\mjb}{mJy~beam$^{-1}$}
\newcommand{\kms}{km~s$^{-1}$}
\newcommand{\tex}{$T_{\rm{ex}}$}

\begin{document}

\title{First Results of an ALMA Band 10 Spectral Line Survey of NGC\,6334I: Detections of Glycolaldehyde (\ce{HC(O)CH2OH}) and a New Compact Bipolar Outflow in HDO and CS}
\author{Brett A. McGuire}
\altaffiliation{B.A.M. is a Hubble Fellow of the National Radio Astronomy\\ Observatory}
\affiliation{National Radio Astronomy Observatory, Charlottesville, VA 22903, USA}
\affiliation{Harvard-Smithsonian Center for Astrophysics, Cambridge, MA 02138, USA}
\author{Crystal L. Brogan}
\affiliation{National Radio Astronomy Observatory, Charlottesville, VA 22903, USA}
\author{Todd R. Hunter}
\affiliation{National Radio Astronomy Observatory, Charlottesville, VA 22903, USA}
\author{Anthony J. Remijan}
\affiliation{National Radio Astronomy Observatory, Charlottesville, VA 22903, USA}
\author{Geoffrey A. Blake}
\affiliation{Division of Geological and Planetary Sciences, California Institute of Technology, Pasadena, CA 91125, USA}
\affiliation{Division of Chemistry and Chemical Engineering, California Institute of Technology, Pasadena, CA 91125, USA}
\author{Andrew M. Burkhardt}
\altaffiliation{Current Address: Submillimeter Array (SMA) Postdoctoral\\ Fellow, Harvard-Smithsonian Center for Astrophysics,\\ Cambridge, MA 02138}
\affiliation{Department of Astronomy, University of Virginia, Charlottesville, VA 22904, USA}
\author{P. Brandon Carroll}
\affiliation{Harvard-Smithsonian Center for Astrophysics, Cambridge, MA 02138, USA}
\author{Ewine F. van Dishoeck}
\affiliation{Leiden Observatory, Leiden University, 2300 RA Leiden, The Netherlands}
\affiliation{Max-Planck Institut f\"{u}r Extraterrestrische Physik, Giessenbachstr. 1, 85748 Garching, Germany}
\author{Robin T. Garrod}
\affiliation{Department of Chemistry, University of Virginia, Charlottesville, VA 22904, USA}
\affiliation{Department of Astronomy, University of Virginia, Charlottesville, VA 22904, USA}
\author{Harold Linnartz}
\affiliation{Sackler Laboratory for Astrophysics, Leiden Observatory, Leiden University, 2300 RA Leiden, The Netherlands}
\author{Christopher N. Shingledecker}
\altaffiliation{Current Address: Alexander von Humboldt\\ Foundation Postdoctoral Research Fellow, Max Planck\\ Institute for Extraterrestrial Physics, Garching,\\ Germany; Institute for Theoretical Chemistry,\\ University of Stuttgart, Stuttgart, Germany}
\affiliation{Department of Chemistry, University of Virginia, Charlottesville, VA 22904, USA}
\author{Eric R. Willis}
\affiliation{Department of Chemistry, University of Virginia, Charlottesville, VA 22904, USA}
\correspondingauthor{Brett A. McGuire}
\email{bmcguire@nrao.edu}

\begin{abstract}

\noindent We present the first results of a pilot program to conduct an ALMA Band~10 spectral line survey of the high-mass star-forming region NGC\,6334I.  The observations were taken in  exceptional weather conditions (0.19~mm precipitable water) with typical system temperatures $T_{\rm{sys}}$~$<$950~K at $\sim$890~GHz.  A bright, bipolar north-south outflow is seen in HDO and CS emission, driven by the embedded massive protostar MM1B.  This has allowed, for the first time, a direct comparison of the thermal water in this outflow to the location of water maser emission from prior 22~GHz VLA observations.  The maser locations are shown to correspond to the sites along the outflow cavity walls where high velocity gas impacts the surrounding material.  We also compare our new observations to prior \emph{Herschel} HIFI spectral line survey data of this field, detecting an order of magnitude more spectral lines (695 vs 65) in the ALMA data.  We focus on the strong detections of the complex organic molecule glycolaldehyde (\ce{HC(O)CH2OH}) in the ALMA data that is not detected in the heavily beam-diluted HIFI spectra.  Finally, we stress the need for dedicated THz laboratory spectroscopy to support and exploit future high-frequency molecular line observations with ALMA.

\end{abstract}

\section{Introduction}
\label{intro}

\begin{figure*}[b!]
    \centering
    \includegraphics[width=0.95\textwidth]{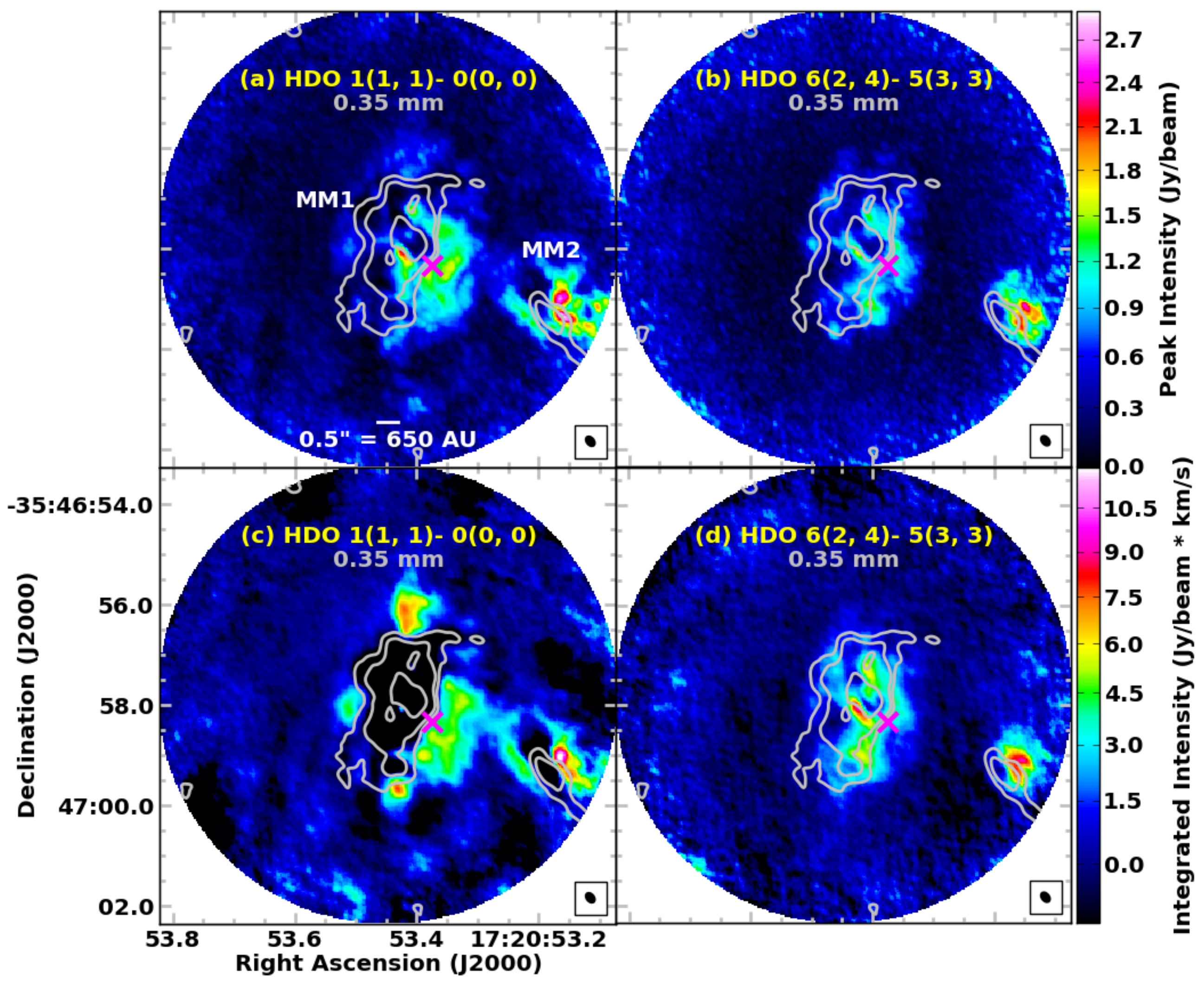}
    \caption{ALMA Band~10 images of the peak (a, b) and integrated intensity (c, d) from $-13.5$ to $+2.5$ \kms\/ for two transitions of HDO with 0.35~mm dust continuum contours overlaid. The 0.35~mm dust continuum contour levels are 50, 75, and 150~K on the Planck temperature scale. Magenta $\times$ symbols show the location of the spectra extracted for MM1 (Fig.~\ref{hifi_comp}). There is substantial absorption of the ground-state transition of HDO toward the MM1 continuum in the integrated intensity maps.  This absorption dominates the integrated intensity map, leaving no net emission signal, whereas the peak intensity maps show the locations of this subsumed HDO emission.  Primary beam correction has been applied with a cutoff at 0.25 of the FWHM. The synthesized beam of $0\farcs23\times 0\farcs16$ (PA$=39\arcdeg$) is shown in the lower right of each panel.}
    \label{map}
\end{figure*}

Observations with the Atacama Large Millimeter Array (ALMA) in Bands 3--7 (84--373~GHz) have proven to be exceptional tools for the detection of new molecular species in the interstellar medium (ISM) and the study of their chemical history and interaction with their physical environment. As a few examples, \citet{Belloche:2014jd} reported the first detection of a branched carbon-chain molecule in the ISM, iso-propyl cyanide (\ce{C3H7CN}), in Band~3 observations of Sgr\,B2.  Later, \citet{McGuire:2017gy} detected methoxymethanol (\ce{CH3OCH2OH}) in surprisingly high abundance toward NGC\,6334I in Band 6~and~7 observations, while \citet{Fayolle:2017bg} identified methyl chloride (\ce{CH3Cl}) for the first time using Band~7 observations of IRAS\,16293-2422.  These observations, among many others, demonstrate the power of ALMA for studies of our molecular universe in the 1--3~mm wavelength range.

Astrochemical observations at higher frequencies, in ALMA Bands 9 (602--720~GHz) and 10 (787--950~GHz) offer complementary benefits to the lower-frequency data, yet few molecular line surveys have been conducted at these frequencies.  Here, we explore two advantages of high-frequency spectral line observations.  First, the fundamental or first few lowest transitions of many small molecules of interest fall into this range.  For example, the HDO $1_{1,1} - 0_{0,0}$ fundamental transition occurs at 893.6~GHz \citep{Messer:1984hh}, providing one of the best opportunities to obtain ground-based measurements of thermal water \citep[see, e.g.,][]{Comito2003}.  

Second, the transitions of most complex organic molecules (COMs) that fall within this frequency range are typically much higher in energy, providing a robust constraint on excitation conditions within a source.  For example, the strongest transitions of glycolaldehyde (\ce{HC(O)CH2OH}), the simplest sugar-related molecule, in Band~6 have upper state energies $E_u$~$\sim$60--200~K in the ground vibrational state.  As a result, analysis of the emission of this molecule can be biased toward lower-excitation conditions, although this can be mitigated through the observation of vibrationally excited states in some cases \citep{Jorgensen:2012dw}.  Complementary observations at higher frequencies, however, provide access to higher-energy lines to rigorously constrain these excitation temperatures -- the strongest transitions of \ce{HC(O)CH2OH} in Band~10 have $E_u$~=~530-630~K -- and can provide needed confirmatory transitions to secure a lower-frequency detection (see, e.g., \citealt{Jorgensen:2012dw}).  

These higher-energy transitions also provide selective access to the warmest molecular gas in a source, which prior studies have shown can have substantially different chemistry from the population probed by the lower-energy transitions accessible at lower frequencies \citep{Crockett:2014er}. There is, however, a relative lack of direct laboratory measurements of molecular spectra above $\sim$600~GHz, meaning that many identifications are made from extrapolated quantum mechanical fits.  For some species this is a reasonably accurate process, but, as will be shown later, the richness of the Band 10 spectra underscores the need for dedicated high-frequency laboratory work.  Finally, at these frequencies it is reasonable to expect that increased dust optical depth effects might ``hide" the deepest, most compact regions of hot molecular cores, and the bright continuum might drive many molecular transitions into absorption.  

To explore the utility of ALMA Band~9/10 observations, we proposed for a full Band~9 survey, and a pilot Band~10 survey, of the high-mass star-forming region NGC\,6334I in Cycle~5.  NGC\,6334I was chosen as the target for three reasons.  First, it is an exceptionally molecular line-rich source \citep{McGuire:2017gy} with a relatively small heliocentric distance of 1.3~kpc \citep{Reid:2014km,Chibueze2014}.  Second, it has previously been targeted by single-dish observations in overlapping frequency ranges by \citet{Zernickel:2012hx} using the Heterodyne Instrument for the Far-Infrared (HIFI; \citealt{deGraauw:2010gy}) on the \emph{Herschel} Space Observatory \citep{Pilbratt:2010en}.  Third, it displays a complex spatial structure consisting of a substantial number of embedded sources and outflows, and several chemically distinct regions separated by only $\sim$2000 au ($\sim$1$\farcs5$; \citealt{Brogan:2016cy,McGuire:2017gy,Bogelund:2018uy}; Figure~\ref{map}). 

Here, we present a first look at ALMA Band~10 observations of any line-rich source, and discuss the results in the context of both probing favorable transitions of light molecules, and in examining the high-excitation lines of complex organic species.  The spectra at these high frequencies are as line-rich as those in the millimeter regime.  Contrary to initial expectations, observations of high-mass star-forming regions like NGC\,6334I at ALMA Band 10 do not appear to be substantially hampered by dust opacity, and are in fact generally better suited than previous single-dish facilities such as Herschel.  These first-look observations demonstrate the power and versatility of high-frequency observations with ALMA.

\section{Observations \& Data Reduction}
\label{observations}

The Band 10 observation occurred 05 April 2018, with 40 ALMA antennas in the array in a nominal C43-3 configuration with a maximum baseline of 532~m providing a 0.21\arcsec~$\times$~0.15\arcsec\/ synthesized beam (robust weighting parameter = 0.5).  The precipitable water vapor at the time of observation was 0.19~mm, and the average resulting system temperature was $T_{\rm{sys}}$~=~926~K at 880~GHz.  Total time on source was 47~minutes, resulting in an RMS noise level of 62~mJy~beam$^{-1}$ in 0.5~\kms\/ channels.  J1517-2422 ($\sim$1.85~Jy at 880~GHz) was used as the flux and bandpass calibration source; J1733-3722 ($\sim$0.43~Jy at 880~GHz) was the phase calibrator.  

At 880~GHz, the half power primary beam width is $\sim$6.6\arcsec\/, which is slightly smaller than the total angular extent of the emitting regions of interest in NGC\,6334I.  We therefore targeted two phase centers to cover the entire source while maximizing the UV coverage and RMS sensitivity in the critical central regions.  As of publication, only the pointing position toward MM1 has been observed; with a phase center of $\alpha$(J2000)~=~17$^{\rm{h}}$20$^{\rm{m}}$53.3$^{\rm{s}}$ $\delta$(J2000)~=~$-35^{\circ}$46\arcmin59.0\arcsec\/ (Figure~\ref{map}). The 0.35~mm continuum was created from
(relatively) line-free channels, and has an aggregate bandwidth and rms noise of 2.7~GHz and 50~\mjb\/, respectively. The method employed for continuum subtraction is described in detail in \citet{Brogan18}. Self-calibration was performed on the continuum and applied to the spectral line data after subtracting the continuum in the uv-plane. The details of the complementary Band~7 data used in this paper are described in \citet{McGuire:2017gy}, while the Band~4 data are described in Appendix~\ref{app:band4}.

For the analysis presented here, spectra were extracted toward MM1 from a single pixel at a position of $\alpha$(J2000) = 17$^{\rm{h}}$20$^{\rm{m}}$53.374$^{\rm{s}}$ $\delta$(J2000)~=~$-35^{\circ}$46\arcmin58.34\arcsec\, (magenta color cross in Figure~\ref{map}).  The location is $\sim$400~au west of the brightest continuum peak, MM1B.  This location was chosen for its proximity to the dense molecular gas while being far enough from the bright continuum that the majority of the molecular lines are not driven into absorption due to the high continuum brightness temperature \citep{McGuire:2017gy}.


The NGC\,6334I region was previously targeted in an extensive, broadband spectral line survey as part of the CHESS (Chemical HErschel Surveys of Star forming regions; \citealt{Ceccarelli:2010hu}) key program using HIFI \citep{Zernickel:2012hx}.  The data used in this manuscript for comparison to our ALMA data were obtained from the \emph{Herschel} Science Archive, ObsID 1342192328.  The HIFI data are used directly as downloaded from the archive, with the exception of the subtraction of a static continuum offset, to place the spectral baseline at $T_A$~=~0~K.



\section{HDO and CS}
\label{HDO}

Most transitions of \ce{H2O} are blocked from the ground by atmospheric absorption features, and thus observations normally rely either on extreme excitation conditions (i.e. maser emission), isotopologues, or space-based observatories to study \ce{H2O}.  Spectra toward NGC~6334I, Orion~KL, and other star-forming regions in the rotational and lowest few ground state transitions of ortho-\ce{H2O} recorded by the {\it Submillimeter Wave Astronomy Satellite (SWAS)} and \emph{Herschel} exhibit broad wing components that arise from heated gas in low and high velocity outflowing gas \citep{Melnick2000,Neufeld2000,vanderTak:2013hk,SanJoseGarcia:2015js}.  Our Band~10 observations provide access to the HDO fundamental $1_{1,1} - 0_{0,0}$ transition at 893.6~GHz ($E_u$~=~43~K) and the higher-energy ($E_u$~=~581~K) $6_{2,4} - 5_{3,3}$ transition at 895.9~GHz \citep{Messer:1984hh}.

Figure~\ref{map} shows the full velocity extent of these HDO transitions as both peak intensity and integrated intensity images.  While no emission signals of this transition were detected in the HIFI data \citep{Emprechtinger2013}, the lower-energy, fundamental HDO transition shows an extended distribution in the high resolution ALMA data. The higher-energy transition is more compact and located closer to the primary continuum sources, likely tracing warmer regions within the source. This pattern is similar to that measured in Orion~KL for two intermediate energy transitions in ALMA Band~6 \citep{Neill2013}. It is notable that the fundamental transition of HDO exhibits significant self-absorption toward the bright continuum sources, and even the $6_{2,4} - 5_{3,3}$ transition is affected by the high continuum opacity and brightness temperature at the continuum peak locations (see Fig.~\ref{map}.)

\begin{figure*}[ht!]
    \centering
    \includegraphics[width=0.95\textwidth]{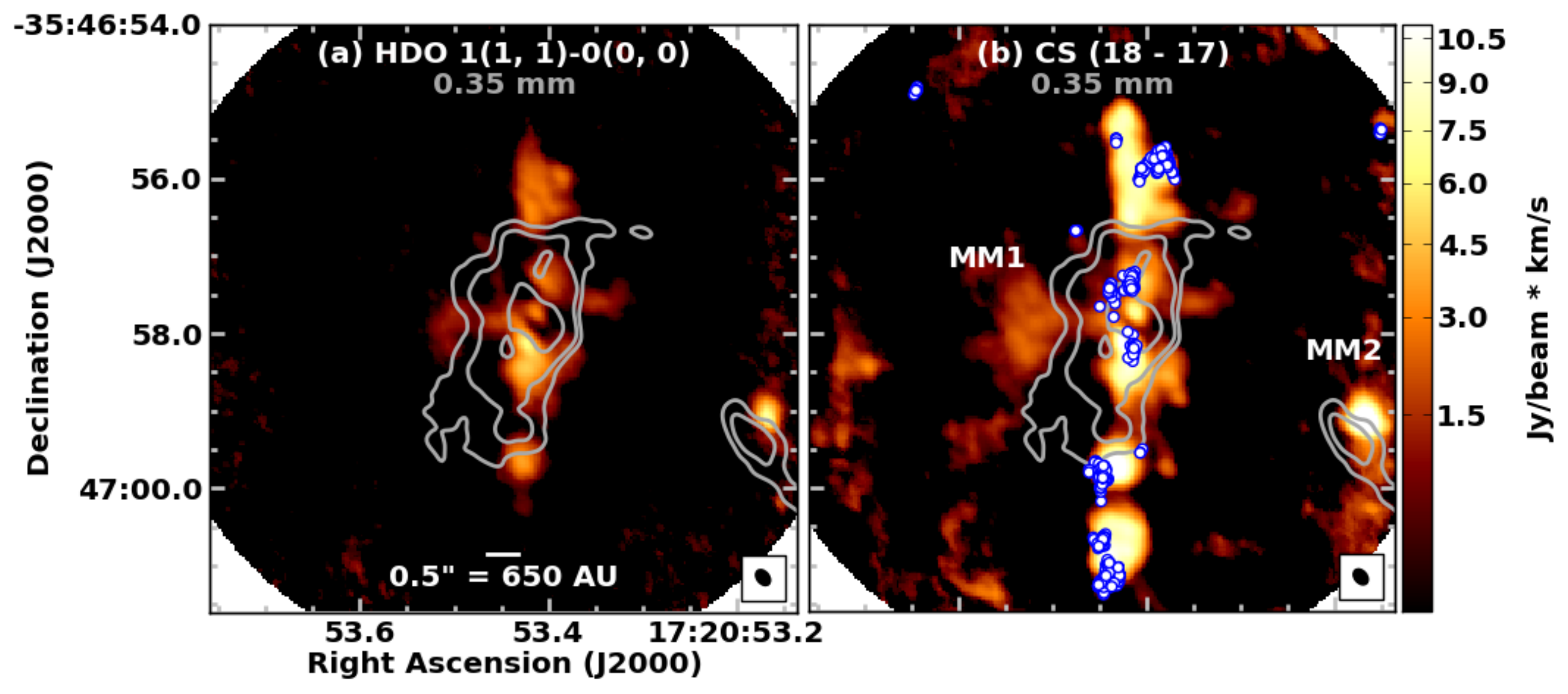}
    \caption{ALMA Band 10 images of the integrated intensity of the redshifted high velocity gas ($-3$ to +2.5 \kms\/; the systemic velocity is $-7$~\kms\/) for the (a) ground state of HDO and the (b) CS (18-17) transition showing a highly collimated north-south outflow emanating from MM1. Blueshifted high velocity emission (not shown) is co-spatial with the redshifted emission, but is much weaker. The 0.35~mm continuum is the same as Fig.~\ref{map}. White circles show the locations of H$_2$O masers detected with the VLA in epoch 2017.8 \citep{Brogan18}. Primary beam correction has been applied with a cutoff at 0.25 the FWHM. The synthesized beam of $0\farcs23\times 0\farcs16$ (PA$=39\arcdeg$) is shown in the lower right of each panel.}
    \label{outflow}
\end{figure*}

Figure~\ref{outflow}a shows only the redshifted, high velocity components of the HDO emission, and reveals a structure that traces a north-south bipolar outflow emanating out of the MM1 continuum peak, centered on the MM1B protostellar source.  Our data also covered the CS $J$~=~18--17 transition ($E_u$~=402~K) at 880.9~GHz \citep{Muller:2005ii}, which shows the same north-south outflow structure as HDO (Figure~\ref{outflow}b). This outflow, called the MM1B N-S outflow has also been detected in ALMA CS (6-5) emission by \citet{Brogan18}. These authors find a total linear extent of $6.2\arcsec$ (8060~au) for MM1B N-S, and that its orientation is nearly in the plane of the sky, leading to a very young dynamical time of only 166 years.  


Extensive observations of \ce{H2O} masers have been made toward NGC\,6334I at cm-wavelengths both with single-dish monitoring \citep{MacLeod:2018hw} and recent complementary Karl G. Jansky Very Large Array (VLA) interferometric observations \citep[][]{Brogan:2016cy,Brogan18}.  By overlaying the locations of cm-wavelength \ce{H2O} masers from the VLA data, it becomes apparent that the masers are predominantly tracing the walls of this outflow cavity where the high velocity gas likely impacts the surrounding quiescent gas.  These locations are consistent with water maser pumping models in which the observed masers arise from velocity coherent structures in the hot gas behind shocks propagating in dense regions \citep{Hollenbach2013,Melnick1993}. The ability to monitor both the thermal water in the outflow and pinpoint the locations of the \ce{H2O} maser emission to the impact sites of the outflow into the surrounding media demonstrates the powerful combination of ALMA Band 10 data with VLA cm-wave observations. In this case, the detection of a collimated outflow in thermal and maser lines along the same axis as the compact radio jet from MM1B \citep{Hunter2018,Brogan18} enables the recognition of these various distinct phenomena as arising from a unified structure.



\subsection{Spectral Line Survey \& Glycolaldehyde (\ce{HC(O)CH2OH})}

The benefits of performing high-frequency spectral line surveys were recognized and exploited by a number of key projects performed with the HIFI instrument aboard \emph{Herschel}.  As mentioned earlier, many small molecules have their first few rotational transitions at these higher frequencies, and surveys with HIFI resulted in the first reported detections of \ce{SH+} \citep{Benz:2010ei}, \ce{HCl+} \citep{DeLuca:2012cv}, \ce{H2O+} \citep{Ossenkopf:2010es}, and \ce{H2Cl+} \citep{Lis:2010ff}.  Additionally, full spectral line surveys provided robust constraints on the excitation conditions and populations of larger molecules ($>$5 atoms) in a variety of extraordinary molecular sources such as Sgr~B2(N) \citep{Neill:2014cb} and Orion-KL \citep{Crockett:2014er}.

A number of spectral line surveys have been carried out toward NGC\,6334I, primarily between 80--270~GHz and by \emph{Herschel} HIFI from 500--1900~GHz (see \citealt{Zernickel:2012hx} and refs. therein).  While these surveys revealed a line-rich source with a complex molecular inventory, the large beam sizes were unable to resolve the underlying structure, and suffered significantly from beam dilution.  The overall angular extent of NGC\,6334I is $\sim$10\arcsec$\times$8\arcsec\/ (13000$\times$10000~au), with most complex molecules emitting over an extent of $\leq$5\arcsec\/ (6500~au) \citep{McGuire:2017gy}.  In a \emph{Herschel} beam of $\sim$25\arcsec\/ at 880~GHz, this mismatch results in a factor of $\gtrsim$25 loss in line brightness due to beam dilution.

\begin{figure*}[t!]
    \centering
    \includegraphics[width=\textwidth]{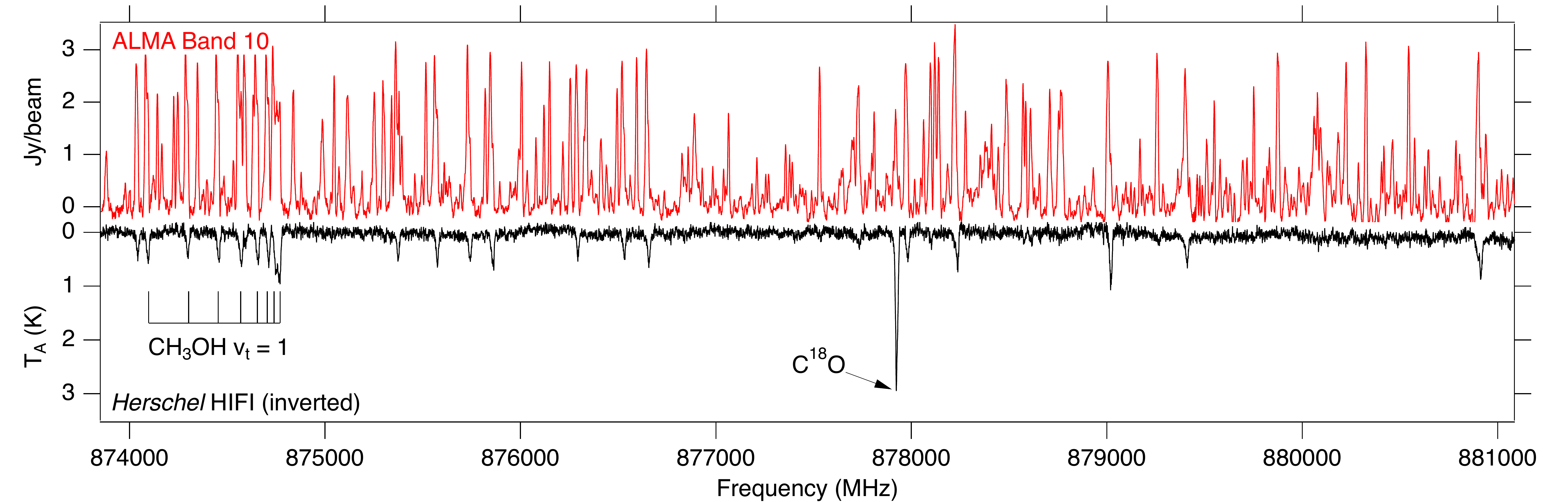}
    \caption{Comparison of this ALMA Band~10 dataset (top; extracted toward MM1) to observations retrieved from the \emph{Herschel} archive and taken as part of the CHESS key program using the HIFI instrument covering the same data range (bottom; \citealt{Zernickel:2012hx}).  The \emph{Herschel} beam ($\sim$25\arcsec~at 880~GHz) covered the entirety of the NGC\,6334I region.  Note that the ALMA spectra are in Jy/beam while the \emph{Herschel} data are in antenna temperature, and that the \emph{Herschel} spectra have been inverted for comparison. A constant offset has been subtracted from the \emph{Herschel} data to remove the continuum for presentation purposes.  Transitions of \ce{CH3OH} $v_t$~=~1 and \ce{C^{18}O} are labeled.}
    \label{hifi_comp}
\end{figure*}

Indeed, when compared to the spectral line survey from HIFI presented in \citet{Zernickel:2012hx}, the most striking feature  of our ALMA Band 10 spectra is the greatly enhanced line density, and the number of lines which are optically thick. Figure~\ref{hifi_comp} shows the full $\sim$8~GHz spectral coverage in the lower sideband of the ALMA data toward MM1 compared with the HIFI data at the same frequency.  The line density in the ALMA spectrum is $\sim$10 times that of HIFI (695 vs 65 lines in the ALMA vs HIFI spectra over the same range). This is exemplified by the emission lines of \ce{CH3OH} ($v_t$~=~1) and \ce{C^{18}O} seen in each spectrum.  The complex molecular emission is significantly enhanced in the ALMA spectra due to its more spatially compact distribution. \ce{C^{18}O}, conversely, is not optically thick in the ALMA data, with a substantially decreased intensity relative to the complex molecules.  This effect is almost certainly due to a more extended distribution which is being partially resolved out in the ALMA observations, but for which the HIFI observations were well-suited.


\begin{figure*}

    \centering
    \includegraphics[width=\textwidth]{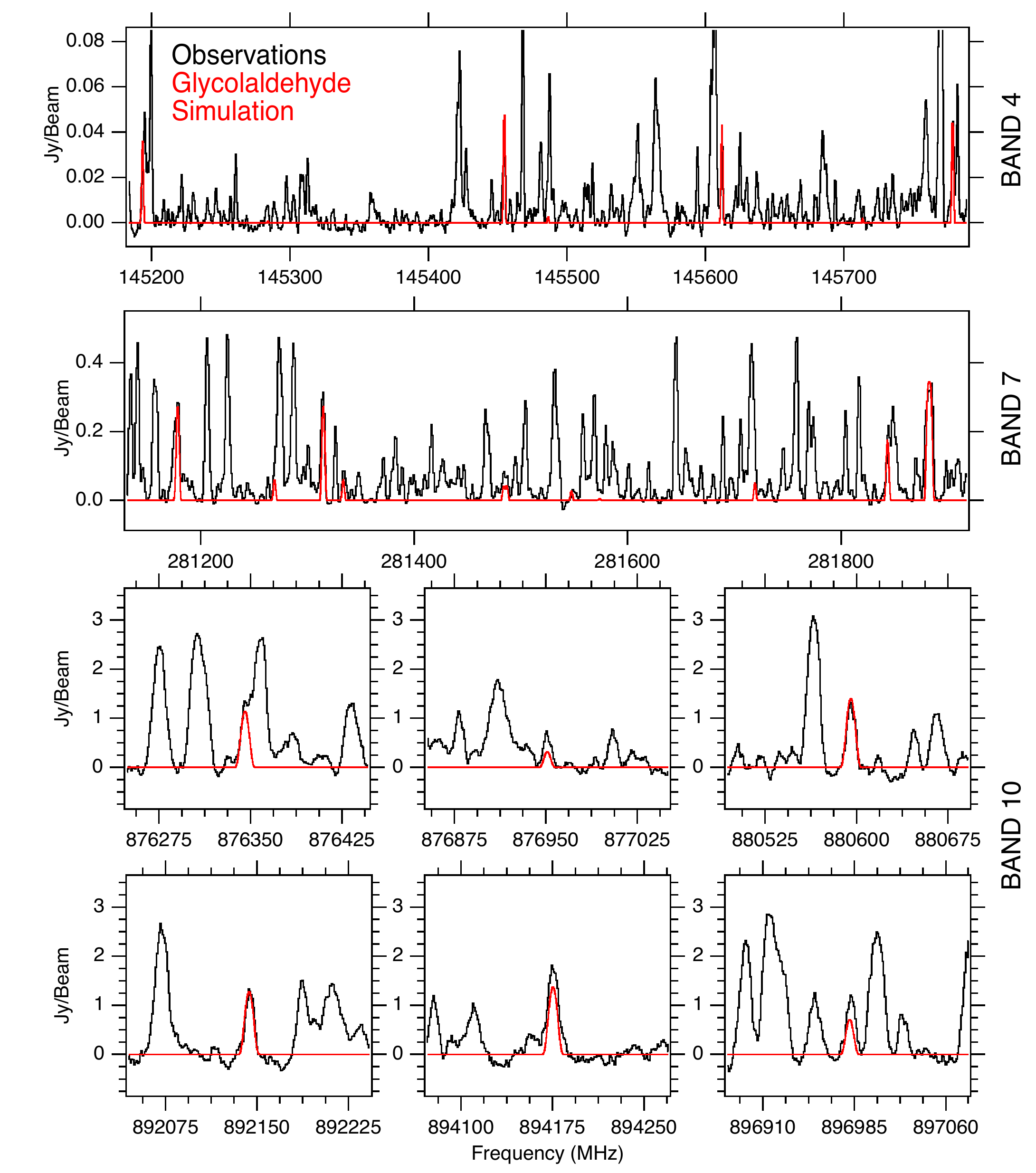}
    \caption{Simulated spectra of \ce{HC(O)CH2OH} plotted in {\color{red}\textbf{red}} over ALMA observations of NGC\,6334I MM1 plotted in \textbf{black}.  The simulated spectra assume $T_{\rm{ex}}$~=~135~K, $\Delta V$~=~3.2~\kms\/, and $N_T$~=~$1.3\times10^{17}$~cm$^{-2}$, with a $v_{lsr}$~=~-7~\kms\/. The ALMA observations have been convolved to a uniform synthesized beam of 0.26\arcsec$\times$0.26\arcsec.  In the lower six panels, smaller regions of the frequency coverage in Band 10 have been selected to show detail. }
    \label{gly}

\end{figure*}


\begin{figure*}[ht!]
    \centering
    \includegraphics[width=0.95\textwidth]{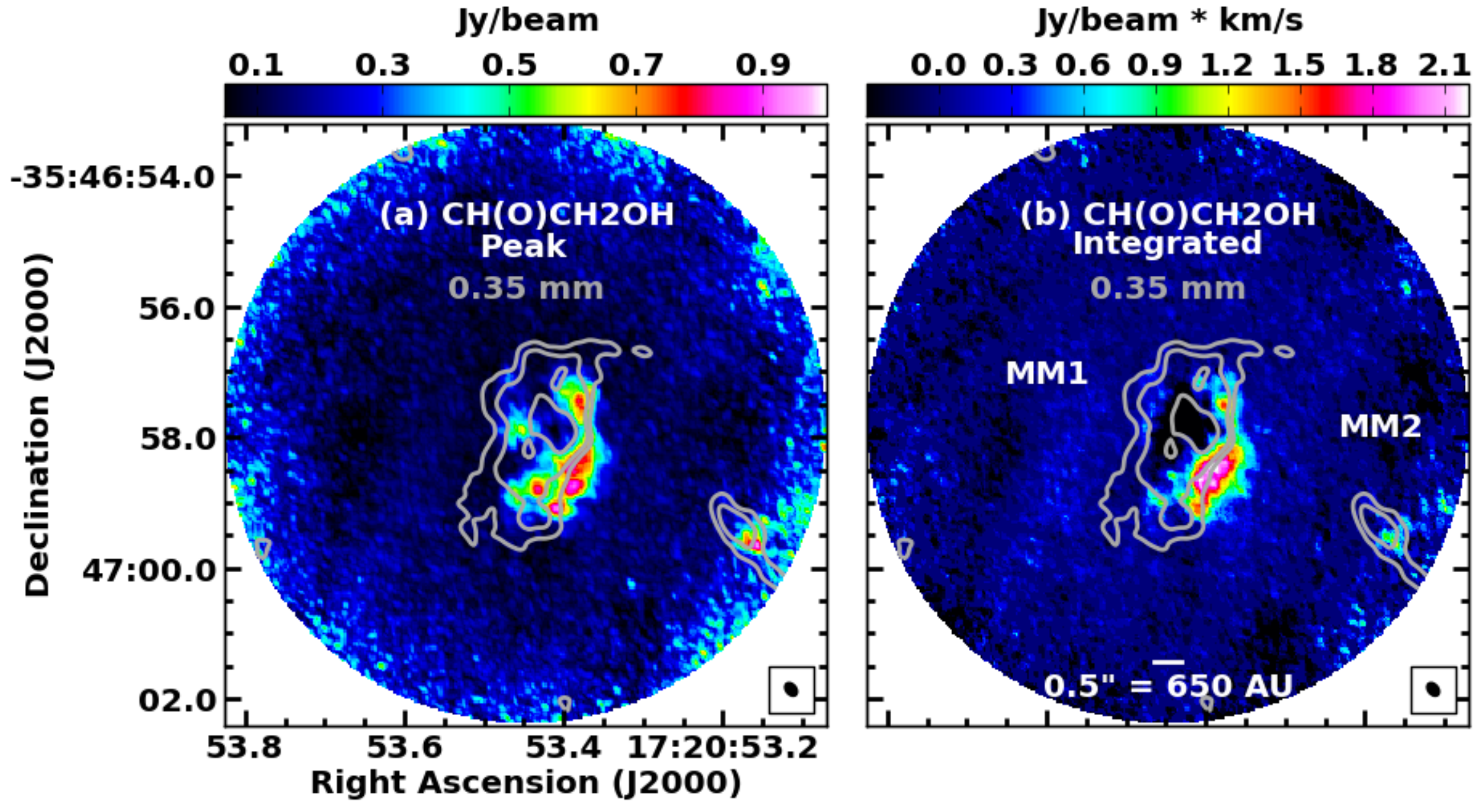}
    \caption{ALMA Band 10 images of the peak (a) and integrated intensity (b) of the \ce{HC(O)CH2OH} transition at 892.12~GHz (28,23 - 27,22; Eu=546~K) shown in Fig.~\ref{gly}. The 0.35~mm continuum is the same as Fig.~\ref{map} and the synthesized beam is shown in the lower right corner. }
    \label{glymap}
\end{figure*}

Because the Band 10 data are so molecular line-rich, they can provide valuable constraints on molecular excitation and column density derivations if they can be robustly analyzed alongside lower-frequency observations.  To test this, we extracted spectra toward the MM1 spectral analysis position from data obtained in previous ALMA observations toward NGC\,6334I in Band~4 (ADS/JAO.ALMA\#2017.1.00661.S) and Band~7 (ADS/JAO.ALMA\#2015.A.00022.T) and in these Band~10 data, all convolved to the smallest common synthesized beam size of 0.26\arcsec$\times$0.26\arcsec.  A full molecular analysis of the Band~10 survey is beyond the scope of this Letter.  Once the full Band~9/10 spectral survey is complete, we plan to provide a fully-reduced line survey to the community.  A preliminary inventory of molecules, however, includes \ce{^{13}CO}, \ce{C^{18}O}, \ce{H2CO}, \ce{HNCO}, \ce{CH3OH}, \ce{^{13}CH3OH}, \ce{CH3^{18}OH}, \ce{CH3OH} $v_t$~=~1, \ce{CH3CN}, \ce{CH2NH}, \ce{CH3NH2}, \ce{NH2CHO}, \ce{CH3CH2OH}, \ce{HC(O)CH2OH}, and \ce{CH3OCHO}.

Here, we focus only on \ce{HC(O)CH2OH}, the simplest sugar-related molecule. This complex organic molecule was successfully detected by ALMA toward the solar-mass protostar IRAS 16293-2422 by \citet{Jorgensen:2012dw}, but has not been reported in the HIFI data at Band 10 frequencies. 

While collisional cross sections for molecules more complex than methanol (\ce{CH3OH}) are generally not available, especially at these frequencies, the high densities in the region (source averaged $n_{\rm{H}}$~$>$10$^6$~cm$^{-3}$; \citealt{Russeil:2010en}) suggest that molecules should be well-described by a single \tex.  We therefore used the formalism described in \citet{Turner:1991um} to simulate the spectrum of \ce{HC(O)CH2OH} across the entire $\sim$750~GHz span in observational coverage, accounting for differences in background continuum temperature at each frequency.  The simulated spectra were converted to Jy/beam from Kelvin using the Planck scale, as the Rayleigh-Jeans approximation introduces significant errors at Band~10 frequencies.   The simulated \ce{HC(O)CH2OH} spectra are enabled by the high-frequency laboratory work of \citet{Carroll:2010gt} that extended the measured frequencies from 354~GHz to 1.2~THz.

Figure~\ref{gly} shows the resulting simulated \ce{HC(O)CH2OH} emission overlaid on the observational data in Bands 4, 7, and 10.  We find that assuming a single excitation temperature ($T_{\rm{ex}}$~=~135~K), linewidth ($\Delta V$~=~3.2~\kms\/), and column density ($N_T$~=~$1.3\times10^{17}$~cm$^{-2}$) across the bands well reproduces the observed emission to zeroth order.  The Band~10 data provide a high-energy anchor to the excitation conditions; the six lines shown in Figure~\ref{gly} have upper state energies between $E_u$~=~530--631~K, compared to $E_u$~=~63--171~K for the Band~4 lines shown.  A full fit of the spectrum, including line contamination and blending from other species, is beyond the scope of this first-look paper. Preliminary work indicates the availability of these high and low-energy anchors provides definite constraints on the excitation temperature of glycolaldehyde to be $\sim$135~K.

The peak and integrated intensities of the \ce{HC(O)CH2OH} transition at 892.12~GHz ($J_{K_a,K_c}$~=~$28_{23,x} - 27_{22,x}$; $E_u$~=~546~K) is shown in Fig.~\ref{glymap}.  Toward the MM1 cluster, these high-energy transitions seem to trace a similar, although not identical, distribution to the high-energy HDO lines shown in Figure~\ref{map}, but has a markedly different distribution toward MM2.  Recent laboratory studies have shown that \ce{HC(O)CH2OH} is readily formed on icy dust grains through hydrogenation of solid CO \citep{Fedoseev:2015ef}.  The observed distribution  may indicate that after being formed in the ice, some of this condense-phased \ce{HC(O)CH2OH}  is being driven into the gas-phase by a non-thermal desorption mechanism such as shock-induced sputtering, as previously hypothesized in observations of complex molecules in galactic center regions \citep{RequenaTorres:2006ki}. Thermal desorption mechanisms are likely also relevant in many warmer regions of the source.

The richness of the Band~10 spectra underscores a need for accurate, high-resolution gas phase spectra of complex molecules from laboratory studies in these frequency ranges.  While laboratory data in the millimeter-wave range below $\sim$500~GHz are increasingly common, the technical difficulties in measuring and analyzing sub-millimeter and THz spectra, combined with a general lack of observational applications, has limited the number of groups working in this regime.  While a dedicated effort was undertaken to provide \ce{HC(O)CH2OH} spectra through 1.2~THz, laboratory spectra for many other complex molecules are missing, and transitions in this range must be extrapolated from lower frequencies, leading to increasingly large uncertainties.

\section{Conclusions}

We have presented a first look at ALMA Band~10 spectral line survey toward a line-rich source -- the high-mass star-forming region NGC\,6334I -- obtained in exceptional weather conditions.  The resulting map shows a bright, bi-polar north-south outflow from the central massive protostar MM1b as traced by both HDO and CS emission.  A comparison to archival \emph{Herschel} HIFI data of the source shows the power of spatially resolving underlying substructure with a beam size well-matched to the source, resulting in the unambiguous identification of \ce{CH(O)CH2OH}.  A wealth of additional transitions suggest the presence of additional complex molecules that can be identified once high resolution laboratory data are available.

\acknowledgements

The authors thank the anonymous referee for their careful evaluation which improved the quality of this manuscript.  This paper makes use of the following ALMA data: ADS/JAO.ALMA\#2017.1.00717.S, \#2017.1.00661.S, and \#2015.A.00022.T. ALMA is a partnership of ESO (representing its member states), NSF (USA) and NINS (Japan), together with NRC (Canada) and NSC and ASIAA (Taiwan) and KASI (Republic of Korea), in cooperation with the Republic of Chile. The Joint ALMA Observatory is operated by ESO, AUI/NRAO and NAOJ.   The National Radio Astronomy Observatory is a facility of the National Science Foundation operated under cooperative agreement by Associated Universities, Inc.  Support for B.A.M. was provided by NASA through Hubble Fellowship grant \#HST-HF2-51396 awarded by the Space Telescope Science Institute, which is operated by the Association of Universities for Research in Astronomy, Inc., for NASA, under contract NAS5-26555. Support for A.M.B. was provided by the NSF through the Grote Reber Fellowship Program administered by Associated Universities, Inc./National Radio Astronomy Observatory and the Virginia Space Grant Consortium. This research made use of NASA’s Astrophysics Data System  Bibliographic  Services,  Astropy,  a community-developed core Python package for Astronomy \citep{astropy}, and APLpy, an open-source plotting package for Python hosted at http://aplpy.github.com.


\appendix

\renewcommand\thefigure{\thesection\arabic{figure}}   
\renewcommand\thetable{\thesection\arabic{table}}    

\setcounter{figure}{0}    
\setcounter{table}{0} 

\twocolumngrid

\section{Band 4 Observations}
\label{app:band4}

The salient parameters for the Cycle 5 Band 4 observations are given in Table~\ref{band4} and will be discussed in detail in a forthcoming paper by Brogan et al.

\begin{deluxetable}{lc}[h!]
\tabletypesize{\footnotesize}
\tablewidth{0pc}
\tablecaption{Observing parameters for Band 4 ALMA data \label{band4}}  
\tablehead{\colhead{Parameter} & \colhead{Band 4 (2.1~mm)}}
\startdata
Project Code			&	ALMA 2017.1.00661.S	\\
Observation date(s) 	&	 2017 Dec 3, 7; 2018, Jan 4 \\
Configuration 	&	 C43-6  	\\
Time on Source (minutes) 	&	170	\\
FWHM Primary Beam 	&	 $0.69\arcmin$ 	\\
Polarization products 	&	 dual linear 	\\
Gain calibrator 	&	 J1713-3418 	\\
Bandpass calibrator 	&	 J1617-5848 	\\
Flux calibrator 	&	  J1617-5848	\\
Spectral window center freqs. (GHz) 	&	 130.5, 131.5, 144.5, 145.4 	\\
Spectral window Bandwidth (MHz)        	&	$4\times 937.5$	\\
Spectral resolution (\kms\/)  & 1.1 \kms\/ \\
Robust parameter  	&	0.5	\\
Ang. Res. ($\arcsec\times\arcsec$ (P.A.$\arcdeg$)) 	& $0.23\times 0.16 (-86.7)$ 	\\
RMS noise per channel (\mjb\/ * \kms)   	& 0.8	
\enddata
\end{deluxetable}

\end{document}